\documentclass[twocolumn,pre,showpacs,showkeys,amsmath,amssymb]{revtex4}
\usepackage{graphicx}
\usepackage{hyperref}

\newcommand{\cb}[1]{\put(#1){\circle*{4}}}
\newcommand{\cw}[1]{\put(#1){\circle{4}}}

\renewcommand{\vec}{}

\begin{document}

\title[Dynamic correlations in lattice gas]
{Dynamic correlations in an ordered c(2$\times$2) lattice gas}

\author{P. Argyrakis}
\email{panos@physics.auth.gr}
\author{M. Maragakis}
\affiliation{Department of Physics, University of Thessaloniki, 54124 Thessaloniki, Greece}
\author{O. Chumak}
\author{A. Zhugayevych}
\email{zhugayevych@iop.kiev.ua}
\affiliation{Institute of Physics, 46 Nauky Prospect, 03650 Kyiv-39, Ukraine}

\date{\today}

\begin{abstract}
We obtain the dynamic correlation function of two-dimensional lattice gas with nearest-neighbor repulsion in ordered c(2$\times$2) phase (antiferromagnetic ordering) under the condition of low concentration of structural defects. It is shown that displacements of defects of the ordered state are responsible for the particle number fluctuations in the probe area. The corresponding set of kinetic equations is derived and solved in linear approximation on the defect concentration. Three types of strongly correlated complex jumps are considered and their contribution to fluctuations is analysed. These are jumps of excess particles, vacancies and flip-flop jumps. The kinetic approach is more general than the one based on diffusion-like equations used in our previous papers. Thus, it becomes possible to adequately describe
correlations of fluctuations at small times, where our previous theory fails to give correct results. Our new analytical results for fluctuations of particle number in the probe area agree well with those obtained by Monte Carlo simulations.
\end{abstract}

% 05.50.+q Lattice theory and statistics (Ising, Potts, etc.)
% 68.43.De Statistical mechanics of adsorbates
% 68.43.Jk Diffusion of adsorbates, kinetics of coarsening and aggregation
% 05.40.-a Fluctuation phenomena, random processes, noise, and Brownian motion

\maketitle

\section{Introduction}

Lattice gas models have been extensively studied (see \cite{_LG}). They are used for the description of various physical phenomena such as surface mass transport, ionic conductivity, etc. and are equivalent to the familiar Ising model with conserved spin dynamics (e.g. Kawasaki dynamics \cite{Kawasaki66}). The general theory of kinetic phenomena in lattice systems is far from complete, despite its long history, started from seminal papers \cite{Kawasaki66,_IsingDyn}. Additional complications and new effects arise in particular models due to nontrivial lattices and complex interactions. All this explains the permanent activity in the field ranging from keystone problems such as the susceptibility of two-dimensional Ising model \cite{Zenine05}, to more specific questions like those considered in our previous studies \cite{Argyrakis05,Chumak1,Chumak2} and other recent papers \cite{Zaluska-Kotur05,Bernardin05}.

The subject of the present paper is kinetic phenomena in two-dimensional lattice gas with nearest neighbor repulsion in the ordered c(2$\times$2) phase (see Fig.~\ref{fig_snapshot1}), which corresponds to an Ising antiferromagnet. Static properties of this system are well studied by means of appropriate approximate or exact methods. Exact expressions for the spatial multipoint correlation functions in grand canonical ensemble at half-filling \cite{_IsingCF} are the most pronounced theoretical results obtained in this field. In contrast, temporal correlation functions are less studied even in the case of thermodynamic equilibrium (excluding the simplest cases like on-site correlation function \cite{Bernardin05}). While most efforts were undertaken to describe critical dynamics, still kinetic phenomena outside the critical region are poorly understood. The standard approach for classical gas based on BBGKY hierarchy is not directly applicable to the case of lattice gas, because the equation for one-particle distribution function involves not only two-particle but also higher distribution functions.

To describe kinetic phenomena, we use the method of essential configurations which can be outlined as follows. There are powerful cluster expansion methods like virial or low-temperature expansions for studying static properties of the ordered equilibrium phase. The main idea of these methods is to reduce the total configuration space to some important (essential) configurations which involve only finite number of structural defects. To study the dynamics of lattice gas, some additional configurations, which are generated by already chosen ones in course of their evolution, should also be considered. These transient configurations, having short living times, do not contribute essentially to static properties but affect the kinetics because of their participation in the displacements of structural defects. Thus, by taking into account mutual transformations of both defect types we may derive kinetic equations in the reduced space of essential configurations. Our main interest is for the two-point two-time correlation function. It will be obtained from the solutions of the mentioned kinetic equations. This function is a building block for macroscopic time correlation functions which play a central role in kinetic phenomena \cite{Balucani03}.

From the point of view of practical application the calculated correlation function can be utilized in theory of adsorbates \cite{Beben03,_Adsorbate}. Modern experimental techniques give the opportunity for detailed investigations of the motion of such small objects as individual atoms on crystal surfaces \cite{_STM} (it is worthwhile to mention that not only kinetics but even femtosecond dynamics can be studied \cite{_UPS}). Thus, calculated correlation functions of particle density fluctuations in scales of one-particle jump may be compared with those experimentally obtained. In such a way some microscopic parameters of the adsorbate-adsorbate and adsorbate-substrate interactions may be obtained from this comparison. In fact, our paper is an attempt to generalize the so-called fluctuative method \cite{Lozano95,Gomer73} to the case of kinetic scales. The correlation function is also connected directly with the dynamic structure factor measurable with diffraction techniques \cite{Chvoj00,Conrad98}. Finally, the theoretical analysis may be used to understand and to interpret data obtained by means of Monte Carlo simulations.

More specifically, the objectives of the paper are as follows: (i) to develop an analytic method for studying kinetic phenomena in ordered lattice systems; (ii) to supplement the approach used in the recent paper \cite{Argyrakis05}; (iii) to explain previous results \cite{Argyrakis05} of MC simulations for short-time correlations of fluctuations in small probe areas.

The paper is organized as follows. In Section~\ref{sec_spec} we specify the system under consideration and define the main quantities. Section~\ref{sec_gen} serves as general outlook to the system: typical snapshots, structural defects, and some insight into kinetic phenomena. Then, in Section~\ref{sec_eq}, the description of essential configurations, derivation and solutions of the kinetic equations for structural defects will be presented. In Section~\ref{sec_cf} the correlation function is calculated. And finally, the comparison of analytical data with the results of MC simulations is given in Section~\ref{sec_MC}.

\section{System specification}  \label{sec_spec}

Consider a lattice gas of $N$ particles hopping on a 2D square lattice of size $L_0\times L_0$ with periodic boundary conditions (i.e., on a torus). Points of the lattice (sites) are denoted by a \emph{single} letter, e.g., $\vec{x}=(x_1,x_2)$, where $x_{1,2}=0,\ldots,L_0-1$. Double occupancy of a site is forbidden. The Hamiltonian is
\begin{equation*} H=\frac{1}{2}\sum_{\vec{x}\neq\vec{y}}U_{\vec{x}\vec{y}}n_{\vec{x}}n_{\vec{y}},
\end{equation*}
where $U_{\vec{x}\vec{y}}=U_1$ if $\vec{x}$ and $\vec{y}$ are nearest neighbors and zero otherwise. The case of repulsion is considered ($U_1>0$). We denote $\beta U_1=\phi$, where $\beta=T^{-1}$ is the inverse temperature. The system is described by two parameters: the average concentration of particles $c$ (in surface science it is usually denoted by $\theta$) and the interaction strength $\phi$.

Also, it is assumed that:

(i) the system is at subcritical temperature, i.e., $\phi>\phi_\text{c}\approx 1.76$ so that $q=e^{-\phi}$ is a small parameter;

(ii) $c$ is nearly 0.5 so that c(2$\times$2) phase is pronounced;

(iii) the system is at thermodynamic equilibrium (which is to imply the absence of domain walls).

Under these conditions the system is represented by two sublattices (Fig.~\ref{fig_snapshot1}): one is almost empty and the other is almost filled. They are distinguished by indices $\vec{\alpha},\vec{\beta},\ldots$ and $\vec{\sigma},\vec{\tau},\ldots$ for the empty and filled sublattices, respectively. The concentrations of excess particles in the empty sublattice and vacancies in the filled one are small enough to treat the system as rarefied gas of structural defects.

The motion of particles is assumed to be realized by instantaneous jumps. Thus, simultaneous displacements of different particles are forbidden (single-particle-jump approximation). The hopping itself is an overcoming of an activation barrier. It should be noted that in real physical systems as in surface adsorbates, all microscopic parameters including adsorption sites, hopping pathways, activation barriers, adatoms interaction, etc. can be calculated by ab initio computations (see e.g. \cite{_abinitio}).

To be specific, we assume the rate of individual particle jumps from a filled site $\vec{x}$ to any unoccupied nearest neighbor site $\vec{y}$ to be given by $W_{\vec{x}\vec{y}}=\nu_0 e^{k\phi}$, where $k$ is the number of filled sites, which are nearest-neighbors to the site $\vec{x}$, and $\nu_0$ is the jump rate in the absence of particle-particle interaction. The detailed balance conditions are satisfied for this choice of jump rate: $W_{\vec{x}\vec{y}}/W_{\vec{y}\vec{x}}=e^{\varepsilon_\vec{x}-\varepsilon_\vec{y}}$, where $\varepsilon_\vec{x}$ is the particle energy in site $\vec{x}$ expressed in units of $T$. Hence in the absence of external perturbations, the system relaxation towards equilibrium is ensured. 

This ``single site energy'' probability, which is widely used in modern surface science, has been suggested in pioneer paper \cite{Bowker78}, in which the results of computer simulations of particle migration for two dimentional lattice gas were reported. The same model was used later to develop analytical approaches in \cite{Chumak80,Reed81}. It is believed now to be a more realistic representation of the diffusive kinetics in experimental systems, although it is slower than the METROPOLIS algorithm.

Without the loss of generality we choose the unit of time in such a way that $\nu_0=1$. Thus the motion of a single particle in the empty lattice are described by the following master equation:
\begin{equation}  \label{Master}
\dot{p}_\vec{x}=-4p_\vec{x}+\sum_\vec{e}p_{\vec{x}+\vec{e}},
\end{equation}
where the sum runs over nearest neighbor sites ($\vec{e}$ means a unit vector). Here and throughout the paper, the length is given in units of a lattice constant.

The main object of the paper is the equilibrium two-point two-time correlation function given by
\begin{equation}  \label{cf}
\left\langle\delta n_\vec{x}(t)\delta n_\vec{y}(0)\right\rangle =\left\langle
n_\vec{x}(t)n_\vec{y}(0)\right\rangle -\left\langle n_\vec{x}(t)\right\rangle\left\langle n_\vec{y}(0)\right\rangle.
\end{equation}
In practical applications other quantity is used: the correlation function of particle number fluctuations in a small probe area, $\left\langle\delta N(t)\delta N(0)\right\rangle$. It can be measured experimentally and useful for surface diagnostics. These two correlation functions are connected by the relation
\begin{equation}  \label{dN2}
\left\langle\delta N(t)\delta N(0)\right\rangle =\sum_{\vec{x},\vec{y}} \left\langle\delta n_\vec{x}(t)\delta n_\vec{y}(0)\right\rangle,
\end{equation}
where the summation is over the probe area. It is more convenient to use ``per site'' quantity, which in case of a square probe area with $L\times L$ size is defined as
\begin{equation}  \label{S}
S_L(t)=\frac{\left\langle\delta N(t)\delta N(0)\right\rangle}{L^2}
\end{equation}
In the limit $L\to\infty$ (in such a way that $L_0\gg L$) we obtain the thermodynamic correlation function $S(t)\equiv S_\infty(t)$, which for $t=0$ can be calculated also by the formula $S(0)=T\left(\partial c/\partial\mu\right)_{T,L_0}$, where $\mu$ is the chemical potential.

The dynamic structure factor can be also expressed via two-point correlation function:
\begin{equation}
S_L(t,\vec{k})=\frac{1}{L^2}\sum_{\vec{x},\vec{y}}\left\langle\delta n_\vec{x}(t)\delta n_\vec{y}(0)\right\rangle e^{-i\vec{k}(\vec{x}-\vec{y})},
\end{equation}
so that $S_L(t,\vec{0})=S_L(t)$ (this normalization differs from that used in \cite{Chvoj00,Conrad98}).

\section{General outlook} \label{sec_gen}

%subsec{Snapshots}

\begin{figure}
\begin{minipage}[b]{1.0\linewidth}
\centering
\includegraphics{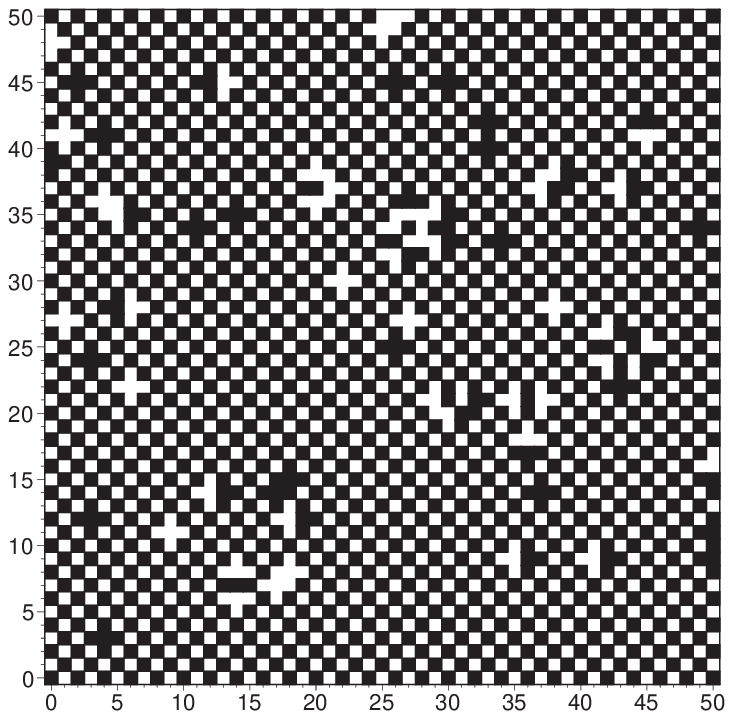}
\caption{A snapshot for $\phi=2.0$ and $c=0.5$. A black square is an excess particle (occupied site), while a white square is a vacancy (empty site).}
\label{fig_snapshot1}
\end{minipage}
\begin{minipage}[b]{1.0\linewidth}
\centering
\includegraphics{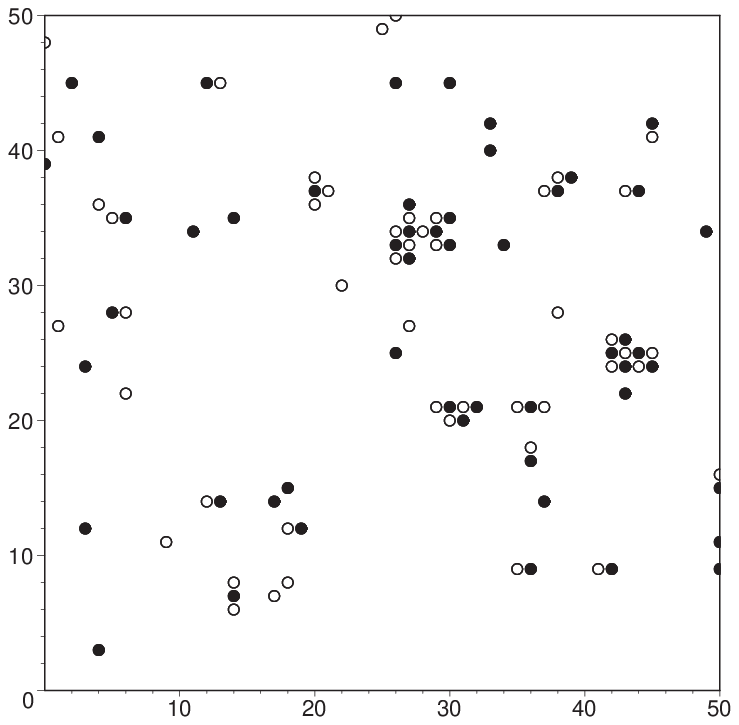}
\caption{``Defects only'' view of Fig.~\ref{fig_snapshot1}. The full circles represent excess particles, while the empty circles are vacancies. The sites which are filled according to the regular ``chessboard'' pattern are not shown.}
\label{fig_snapshot2}
\end{minipage}
\begin{minipage}[b]{1.0\linewidth}
\centering
\includegraphics{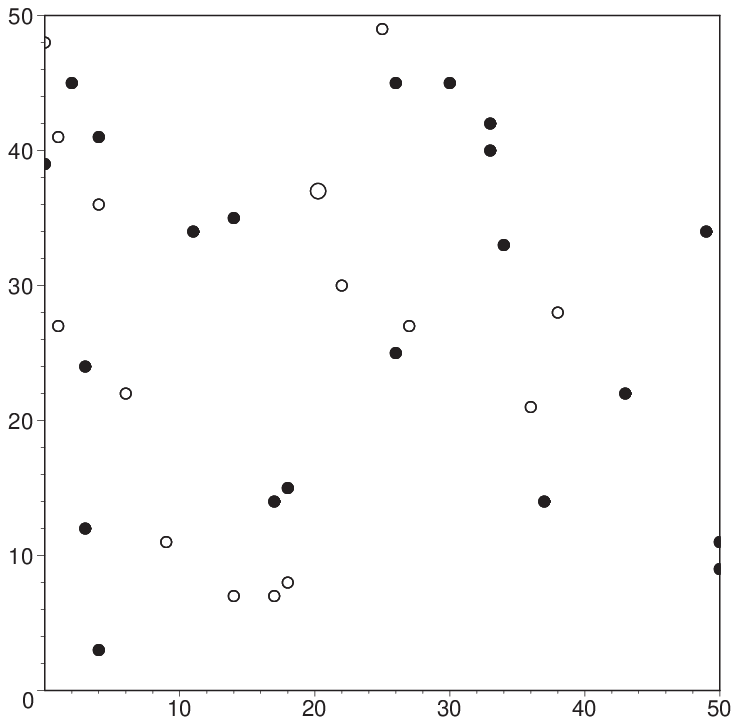}
\caption{Snapshot of the topological charges after the subtraction of the excess particle -- vacancy pairs. The large circle in the upper middle of the picture is a dimer of vacancies.}
\label{fig_snapshot3}
\end{minipage}
\end{figure}

A typical snapshot (obtained in the course of MC simulations) is shown in Fig.~\ref{fig_snapshot1}. The ``chessboard'' phase is clearly seen, even for $\phi$ close to its critical value. The ``defects only'' view is presented in Fig.~\ref{fig_snapshot2}, where only excess particles in the empty sublattice and vacancies in the filled one are shown. This view supports the idea that we can treat the system as rarefied gas of structural defects. To elucidate the nature of these defects we redraw the snapshots, leaving only the topological charge. This quantity is defined as the difference between the number of excess particles and vacancies belonging to the same cluster of connected defects in Fig.~\ref{fig_snapshot2}. To produce Fig.~\ref{fig_snapshot3} we define a cluster as a group of defects which are nearest neighbors, and place the topological charges at the mean coordinates of the clusters. The result is shown in Fig.~\ref{fig_snapshot3}, from where it is clearly seen that the seeming variety of the structural defects in Fig.~\ref{fig_snapshot2} actually represents the fluctuating deformations of some basic defects of charges $1$, $-1$, and $0$. Also sparsely populated pairs of defects (the large circle in Fig.~\ref{fig_snapshot3}) are present.

Before classifying these basic structural defects, it should be noted that in a real system domain walls are often presented as nonequilibrium long-living metastable configurations. Nuclei of such inclusions are always present as large continuous areas of defects, observed in Fig.~\ref{fig_snapshot2}. The true domain walls make the whole system essentially inhomogeneous, lacking good ergodic properties. Therefore, we assume that there are no domain walls in the neighborhood of the probe area.

%subsec{Structural defects}

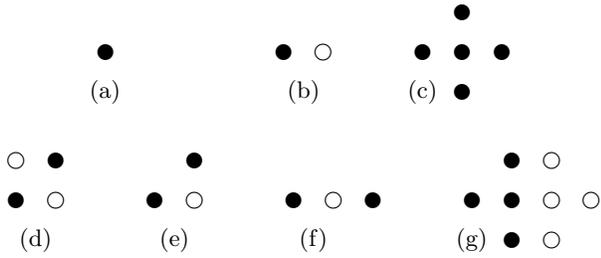
\begin{figure}[h]
\unitlength=1.5pt
\centering
\begin{picture}(100,20)
\put(0,10){\begin{picture}(0,0)
  \put(0,-10){\makebox(0,0)[cc]{(a)}}
  \cb{0,0}
  \end{picture}} 
\put(45,10){\begin{picture}(0,0)
  \put(5,-10){\makebox(0,0)[cc]{(b)}}
  \cb{0,0}\cw{10,0}
  \end{picture}} 
\put(90,10){\begin{picture}(0,0)
  \put(-10,-10){\makebox(0,0)[cc]{(c)}}
  \cb{0,0}\cb{10,0}\cb{0,10}\cb{-10,0}\cb{0,-10}
  \end{picture}}
\end{picture}
\\ \vspace{25pt}
\begin{picture}(145,20)
\put(0,10){\begin{picture}(0,0)
  \put(5,-10){\makebox(0,0)[cc]{(d)}}
  \cb{0,0}\cw{10,0}\cw{0,10}\cb{10,10}
  \end{picture}}
\put(35,10){\begin{picture}(0,0)
  \put(5,-10){\makebox(0,0)[cc]{(e)}}
  \cb{0,0}\cw{10,0}\cb{10,10}
  \end{picture}}
\put(70,10){\begin{picture}(0,0)
  \put(5,-10){\makebox(0,0)[cc]{(f)}}
  \cb{0,0}\cw{10,0}\cb{20,0}
  \end{picture}}
\put(125,10){\begin{picture}(0,0)
  \put(-10,-10){\makebox(0,0)[cc]{(g)}}
  \cb{0,0}\cw{10,0}\cb{-10,0}\cb{0,10}\cb{0,-10}\cw{20,0}\cw{10,10}\cw{10,-10}
  \end{picture}}
\end{picture}
\caption{All structural defects up to $q^4$ level: (a) excess particle, (b) flip-flop pair, (c) excess particle monomer, (d) flip-flop tetrad, (e,f) transient configurations during monomer jump, (g) isolated flip-flop pair. Only essential sites are shown, positions with respect to sublattices can be easily guessed. Configurations obtained by color inversion must be also included.}
\label{fig_StrDef}
\end{figure}

Various structural defects may be classified by their concentrations at exact half-filling. To get rid of possible ambiguity in defects definitions, we use the following rule: two structurally similar defects are called distinguishable at level $q^n$ if their concentrations at exact half-filling differs by $O(q^k),\ k\leq n$. To make the ground for the study of kinetic phenomena we will consider the lowest-order defects shown in Fig.~\ref{fig_StrDef} up to level $q^4$. It is worthwhile to note that in Fig.~\ref{fig_snapshot2} (where $q=0.8q_c$) only five clusters have smaller concentrations: $q^5$, $q^5$, $q^6$, $q^7$, $q^{13}$.

First of all, excess particles and vacancies can be easily distinguished as alone circles in Fig.~\ref{fig_snapshot2}. The concentration of both defect types at exact half-filling is given by \cite{_LG}
\begin{equation}
n=v=\frac{1}{2}\left(1-\sqrt[8]{1-\frac{16q^2}{(1-q)^4}}\right) \approx q^2+4q^3
\end{equation}
(for simple cubic lattice $n=v\approx q^3$ \cite{Chumak2}). Only these two structural defects constitute the level $q^2$ because all other defects either have concentration $o(q^2)$ or are indistinguishable from these two at the level $q^2$.

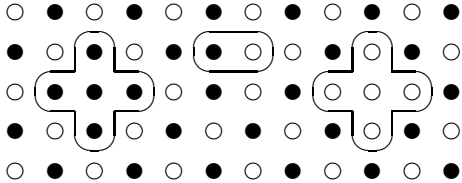
\begin{figure}[h]
\unitlength=1.5pt
\centering
\begin{picture}(110,45)(0,0)
\cw{0,40}\cb{10,40}\cw{20,40}\cb{30,40}\cw{40,40}\cb{50,40}\cw{60,40}\cb{70,40}\cw{80,40}\cb{90,40}\cw{100,40}\cb{110,40}
\cb{0,30}\cw{10,30}\cb{20,30}\cw{30,30}\cb{40,30}\cb{50,30}\cw{60,30}\cw{70,30}\cb{80,30}\cw{90,30}\cb{100,30}\cw{110,30}
\cw{0,20}\cb{10,20}\cb{20,20}\cb{30,20}\cw{40,20}\cb{50,20}\cw{60,20}\cb{70,20}\cw{80,20}\cw{90,20}\cw{100,20}\cb{110,20}
\cb{0,10}\cw{10,10}\cb{20,10}\cw{30,10}\cb{40,10}\cw{50,10}\cb{60,10}\cw{70,10}\cb{80,10}\cw{90,10}\cb{100,10}\cw{110,10}
\cw{0, 0}\cb{10, 0}\cw{20, 0}\cb{30, 0}\cw{40, 0}\cb{50, 0}\cw{60, 0}\cb{70, 0}\cw{80, 0}\cb{90, 0}\cw{100, 0}\cb{110, 0}
\put(15,20){\oval(20,10)[l]}\put(25,20){\oval(20,10)[r]}\put(20,15){\oval(10,20)[b]}\put(20,25){\oval(10,20)[t]}
\put(55,30){\oval(20,10)}
\put(85,20){\oval(20,10)[l]}\put(95,20){\oval(20,10)[r]}\put(90,15){\oval(10,20)[b]}\put(90,25){\oval(10,20)[t]}
\end{picture}
\caption{Structural defects at $q^3$ level, from left to right: excess particle monomer, flip-flop pair, vacancy monomer. Essential sites are encircled.}
\label{fig_StrDef_q3}
\end{figure}

At the level $q^3$ of classification we distinguish excess particle surrounded by only occupied sites and excess particle with vacancies in the nearest neighborhood (see Fig.~\ref{fig_StrDef_q3}). In the first case the structural defect is called excess particle monomer, Fig.~\ref{fig_StrDef}(c). In the second case only one vacancy is allowed to have concentration distinguishable from zero at level $q^3$. This is so called ``flip-flop'' pair, Fig.~\ref{fig_StrDef}(b), that is a pair of adjacent excess particle and vacancy. Their concentration per site of a sublattice is
\begin{equation}  \label{n_ff}
2\left\langle n_\vec{0} v_\vec{e}\right\rangle \approx 2q^3.
\end{equation}
Note that this value is considerably greater than in the absence of correlation, $2nv\approx q^4$. The concentration of monomers is
\begin{equation}  \label{n_mono}
\left\langle n_\vec{0}\prod_\vec{e}(1-v_\vec{e})\right\rangle =\left\langle
n_\vec{0}\right\rangle -\sum_\vec{e}\left\langle n_\vec{0} v_\vec{e}\right\rangle+\ldots \approx q^2-5q^4.
\end{equation}
The difference between excess particle, Fig.~\ref{fig_StrDef}(a), and its monomer, Fig.~\ref{fig_StrDef}(c) can be illustrated by the equation
\begin{equation*}
\unitlength=0.5pt
\left(\mbox{\begin{picture}(24,20)(-12,-5) \cb{0,0} \end{picture}}\right)
=  \left(\mbox{\begin{picture}(24,20)(-12,-5) \cb{0,0}\cb{10,0}\cb{0,10}\cb{-10,0}\cb{0,-10} \end{picture}}\right)
+4\!\left(\mbox{\begin{picture}(24,20)(-12,-5) \cb{0,0}\cw{10,0}\cb{0,10}\cb{-10,0}\cb{0,-10} \end{picture}}\right)
+4\!\left(\mbox{\begin{picture}(24,20)(-12,-5) \cb{0,0}\cw{10,0}\cw{0,10}\cb{-10,0}\cb{0,-10} \end{picture}}\right)
+2\!\left(\mbox{\begin{picture}(24,20)(-12,-5) \cb{0,0}\cw{10,0}\cb{0,10}\cw{-10,0}\cb{0,-10} \end{picture}}\right)
+4\!\left(\mbox{\begin{picture}(24,20)(-12,-5) \cb{0,0}\cw{10,0}\cw{0,10}\cw{-10,0}\cb{0,-10} \end{picture}}\right)
+  \left(\mbox{\begin{picture}(24,20)(-12,-5) \cb{0,0}\cw{10,0}\cw{0,10}\cw{-10,0}\cw{0,-10} \end{picture}}\right),
\end{equation*}
from which also the notion of ``essential sites'' used in the figures can be understood. Only monomers and ``flip-flop'' pairs constitute the level $q^3$.

The structural defects of the level $q^4$ are monomers, isolated flip-flop pairs, flip-flop tetrads, and transient configurations arising in the course of monomer jumps (Fig.~\ref{fig_StrDef}(c-g), see also Fig.~\ref{fig_confE}). Despite the low concentration of transient configurations ($\sim q^4$), they are innegligible for description of monomer motion. Starting from this level one must consider also monomer pairs, that is two-particle configurations. At $q^4$-level this reduces to taking into account onsite generation-recombination of vacancy and excess particle monomers and onsite exclusion principle for identical monomers.

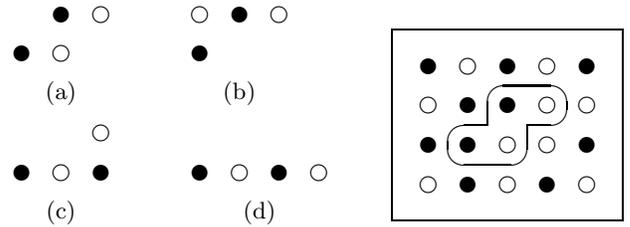
\begin{figure}[h]
\unitlength=1.5pt
\centering
\begin{picture}(85,55)
\put(0,40){\begin{picture}(0,0)
  \put(10,-10){\makebox(0,0)[cc]{(a)}}
  \cb{0,0}\cw{10,0}\cb{10,10}\cw{20,10}
  \end{picture}}
\put(45,40){\begin{picture}(0,0)
  \put(10,-10){\makebox(0,0)[cc]{(b)}}
  \cb{0,0}\cw{0,10}\cb{10,10}\cw{20,10}
  \end{picture}}
\put(0,10){\begin{picture}(0,0)
  \put(10,-10){\makebox(0,0)[cc]{(c)}}
  \cb{0,0}\cw{10,0}\cb{20,0}\cw{20,10}
  \end{picture}}
\put(45,10){\begin{picture}(0,0)
  \put(15,-10){\makebox(0,0)[cc]{(d)}}
  \cb{0,0}\cw{10,0}\cb{20,0}\cw{30,0}
  \end{picture}}
\end{picture}
\quad
\fbox{
\begin{picture}(52,44)(-5,-7)
 \cb{0,30}\cw{10,30}\cb{20,30}\cw{30,30}\cb{40,30}
 \cw{0,20}\cb{10,20}\cb{20,20}\cw{30,20}\cw{40,20}
 \cb{0,10}\cb{10,10}\cw{20,10}\cw{30,10}\cb{40,10}
 \cw{0, 0}\cb{10, 0}\cw{20, 0}\cb{30, 0}\cw{40, 0}
 \put(15,10){\oval(20,10)[l]} \put(25,20){\oval(20,10)[r]}
 \put(15,10){\oval(20,10)[br]}\put(25,20){\oval(20,10)[tl]}
 \put(25,10){\line(0,1){5}}\put(15,15){\line(0,1){5}}
\end{picture}}
\caption{Generation-recombination precursors. The structural defect (a) is also depicted in the box to show its position with respect to sublattices.}
\label{fig_StrDef_GR}
\end{figure}

The generation-recombination processes must be considered for understanding the long-time asymptotics of the correlation function \cite{Argyrakis05,Chumak1,Chumak2}). The generation passes in three steps. The first two are the creation of a double flip-flop pair resulting in one of the transient configurations shown in Fig.~\ref{fig_StrDef_GR} (their concentration is $\sim q^5$). At the last step adjacent vacancy and excess particle monomers are created.

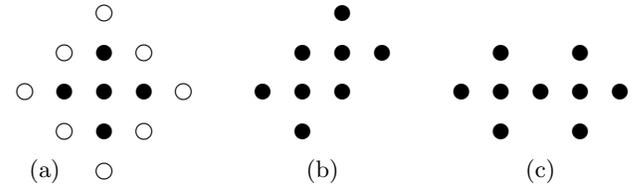
\begin{figure}[h]
\unitlength=1.5pt
\centering
\begin{picture}(150,40)
\put(20,20){\begin{picture}(0,0)
  \put(-15,-20){\makebox(0,0)[cc]{(a)}}
  \cb{0,0}\cb{10,0}\cb{0,10}\cb{-10,0}\cb{0,-10}
  \cw{20,0}\cw{10,10}\cw{0,20}\cw{-10,10}\cw{-20,0}\cw{-10,-10}\cw{0,-20}\cw{10,-10}
  \end{picture}}
\put(70,20){\begin{picture}(0,0)
  \put(5,-20){\makebox(0,0)[cc]{(b)}}
  \cb{0,0}\cb{10,0}\cb{0,10}\cb{-10,0}\cb{0,-10}\cb{10,10}\cb{20,10}\cb{10,20}
  \end{picture}}
\put(120,20){\begin{picture}(0,0)
  \put(10,-20){\makebox(0,0)[cc]{(c)}}
  \cb{0,0}\cb{10,0}\cb{0,10}\cb{-10,0}\cb{0,-10}\cb{20,0}\cb{20,10}\cb{30,0}\cb{20,-10}
  \end{picture}}
\end{picture}
\caption{(a) Isolated monomer, (b) side and (c) corner dimers of excess particle.}
\label{fig_StrDef_dimer}
\end{figure}

At the level $q^6$ isolated monomers become distinguishable from side and corner dimers shown in Fig.~\ref{fig_StrDef_dimer}. The concentrations of these structural defects have leading terms of the low-temperature expansion $q^2-13q^4$, $q^4-8q^6$, and $q^4-9q^6$, respectively. At this level the short-range attraction between monomers arises, that can be observed as small excess in concentration of dimers compared to squared concentration of monomers, which is $\approx q^4-10q^6$. The number of dimer configurations increases when $c\neq0.5$ (as squared monomer concentration) and in that case they may contribute significantly to mass transport \cite{Argyrakis05,Chumak1,Chumak2}.

\medskip

\begin{figure}
\centering
\includegraphics{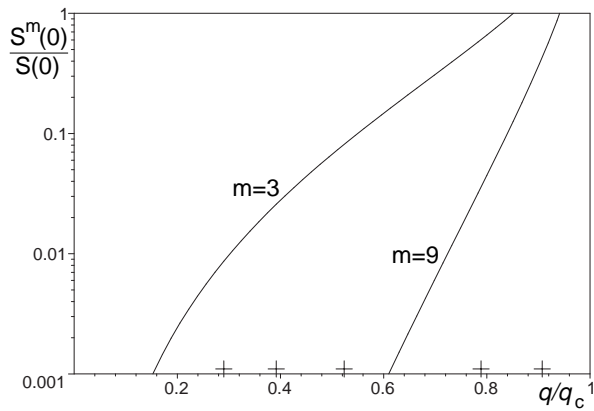}
\caption{Relative contribution of long-distance components of the correlation function to $S(0)$ at exact half-filling according to Eq.~\ref{S^l} depending on interaction strength $q/q_\text{c}\equiv e^{\phi_\text{c}-\phi}$. Parameters of MC simulations (see Sec.~\ref{sec_MC}) are marked by crosses on the horizontal axis $\phi=\{3.0,2.7,2.4,2.0,1.86\}$ (from left to right).} \label{fig_cfat0}
\end{figure}

Obviously, when approaching the critical point or moving away from exact half-filling the larger structural defects should be accounted to obtain reliable results. To understand the importance of large defects for lattice gas statistics we calculate their contribution to $S(0)$, using the analytic expression for $\left\langle\delta n_\vec{x}\delta n_\vec{0}\right\rangle =\left\langle\delta n_\vec{x}(0)\delta n_\vec{0}(0)\right\rangle$ at exact half filling (the expression is not explicit but can be derived by recurrent procedure \cite{_IsingCF}). This can be done by evaluating
\begin{equation}  \label{S^l}
S^m(0)=\sum_{|x_1+x_2|\geq m}\left\langle\delta n_{(x_1x_2)}\delta n_{(00)}\right\rangle
\end{equation}
and comparing it with $S(0)\equiv S^0(0)$. The result is shown in Fig.~\ref{fig_cfat0} for $m=3$ and 9. Roughly speaking, the upper curve, in which short-distance correlations ($|x_1+x_2|\leq2$) are excluded, gives an estimate of an error arising when only monomers and flip-flop pairs are taken into account. This means that the contribution of long-distance correlations in the case of large values of $\phi$ is negligible. Thus, our approach seems quite reliable when $\phi\gtrsim 2.5$ ($T\lesssim 0.7\,T_c$).

%subsec{Kinetics}

For further analysis it is useful to distinguish two time scales of the kinetic phenomena. At small time scale we have a diversity of structural defects (typical snapshot is shown in Fig.~\ref{fig_snapshot1}). The major part of them have very short lifetime. Their fast evolution is due to strong repulsive interaction with the nearest neighbors. The contribution of short-term fluctuations to the dynamic correlation function decays exponentially with time. At long times only topologically stable structural defects (monomers of excess particles and vacancies, as well as domain walls) survive and contribute to the correlation function. The motion of
long-living defects results in such macroscopic phenomena as diffusion, segregation, domain growth. A typical decay of the corresponding correlation function is given by a power-law form. In the next section we will derive kinetic equations governing the evolution of structural defects.

\section{Kinetic equation for structural defects} \label{sec_eq}

%subsec{A method of essential configurations}

The configuration space of the system is extremely large, $2^{L_0^2}$. But observations made in the previous section together with general principles of statistical physics suggest that only a tiny part of the entire configuration space is actually occupied. In the present paper we derive the kinetic equation in the reduced space of \emph{essential configurations}. The general sketch of the method is as follows. At first we set the desired concentration level, say $q^k$, and choose configurations with concentration no less than $q^k$ (note that each configuration has $L_0^2$ translationally invariant copies, therefore we use ``per site'' quantities). Thus, we obtain the configuration space of structural defects. Then we determine transient configurations which may be considered as ``deformations'' of the structural
defects in the course of their displacements. It should be emphasized that the search of the transient states is not a specific problem of lattice systems exclusively. It is a general problem of kinetics. And finally, we write the standard master equation for the Markov chain in the reduced space of essential configurations (defects + their deformations). We have to solve this set of equations up to order $q^k$. If we are interested in the correlation function we may exclude transient configurations (with concentration $o(q^k)$) from this set. The equations obtained in such a way govern the evolution of structural defects only. This evolution is nonmarkovian and this is the main effect of transient states.

In our case the first nonvanishing level is $q^2$, which includes monomers only. We will also consider the next level, $q^3$, that is, flip-flop pairs in order to look for the corrections due to higher orders defects.

%subsec{Kinetic equation for monomers}

The motion of monomers was explained in \cite{Chumak1}. The main idea of that paper may be outlined as follows. Let us consider an excess atom shown in Fig.~\ref{fig_moveE}. In such configuration the most frequent processes are jumps like that from D$_1$ to an empty site with rate $e^{\phi}$. Let it will be site C. The new configuration is short-lived because the most probable process has rate $e^{3\phi}$. Now there are two possibilities. The first is that the displaced particle returns to site D$_1$ so that the final state coincides with the initial one (i.e. $\text{D}_1\to\text{C}\to\text{D}_1$). The second is that the particle in site A jumps to site D$_1$ and the final state is one with the excess particle displaced from A to C (i.e. $\text{D}_1\to\text{C}$, $\text{A}\to\text{D}_1$). Both possibilities have equal probabilities. Therefore if one neglects the living time of the transient state, the rate of this complex (two-step) jump of the defect will be given by $\frac{1}{2}e^\phi$ \cite{Chumak1}. Similarly the rate for $\text{A}\to\text{B}$ complex jump is $2\times\frac{1}{2}e^\phi$, where the factor of two arises from two possible transient states: D$_1$ and D$_2$. The same reasoning for vacancy displacements (Fig.~\ref{fig_moveV}) results in rate value equal to 1 for $\text{A}\to\text{B}$ jump and $\frac{1}{2}$ for $\text{A}\to\text{C}$ jump.

\unitlength=3pt
\begin{figure}[h]
\begin{minipage}[b]{1.0\linewidth}
\centering
\begin{picture}(40,35)
\multiput(0,0)(0,20){2}{
  \multiput( 0, 0)(20,0){3}{\circle{4}}
  \multiput(10,10)(20,0){2}{\circle{4}}}
\multiput(0,0)(0,20){2}{
  \multiput(10, 0)(20,0){2}{\circle*{4}}
  \multiput( 0,10)(20,0){3}{\circle*{4}}}
\put(10,10){\circle*{4}} \put(11, 6){\scriptsize{A}} \put(21,16){\scriptsize{B}}
\put(31,6){\scriptsize{C}} \put(21, 6){\scriptsize{D$_1$}} \put(11,16){\scriptsize{D$_2$}}
\put(10,10){\vector(1,1){8.5}} \put(10,10){\vector(1,0){18}}
\end{picture}
\caption{Displacement of excess particle monomer.}
\label{fig_moveE}
\end{minipage}
\begin{minipage}[b]{1.0\linewidth}
\centering
\begin{picture}(40,35)
\multiput(0,0)(0,20){2}{
  \multiput( 0, 0)(20,0){3}{\circle*{4}}}
\multiput(0,0)(0,20){2}{
  \multiput(10, 0)(20,0){2}{\circle{4}}
  \multiput( 0,10)(20,0){3}{\circle{4}}}
\put(10,10){\circle{4}} \put(30,10){\circle*{4}} \put(10,30){\circle*{4}}
\put(30,30){\circle*{4}} \put(11, 6){\scriptsize{A}} \put(21,16){\scriptsize{B}}
\put(31,6){\scriptsize{C}} \put(21, 6){\scriptsize{D$_1$}} \put(11,16){\scriptsize{D$_2$}}
\put(10,10){\vector(1,1){8.5}} \put(10,10){\vector(1,0){18}}
\end{picture}
\caption{Displacement of vacancy monomer.} \label{fig_moveV}
\end{minipage}
\end{figure}

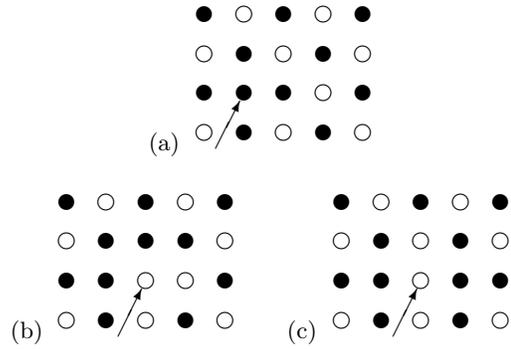
\begin{figure}[h]
\unitlength=1.5pt
\centering
\begin{picture}(55,40)(-10,-5)
\put(-10,-3){\makebox(0,0)[cc]{(a)}}
\cb{0,30}\cw{10,30}\cb{20,30}\cw{30,30}\cb{40,30}
\cw{0,20}\cb{10,20}\cw{20,20}\cb{30,20}\cw{40,20}
\cb{0,10}\cb{10,10}\cb{20,10}\cw{30,10}\cb{40,10}
\cw{0, 0}\cb{10, 0}\cw{20, 0}\cb{30, 0}\cw{40, 0}
\put(3,-4){\vector(1,2){6}}
\end{picture}
\\ \vspace{10pt}
\begin{picture}(55,40)(-10,-5)
\put(-10,-3){\makebox(0,0)[cc]{(b)}}
\cb{0,30}\cw{10,30}\cb{20,30}\cw{30,30}\cb{40,30}
\cw{0,20}\cb{10,20}\cb{20,20}\cb{30,20}\cw{40,20}
\cb{0,10}\cb{10,10}\cw{20,10}\cw{30,10}\cb{40,10}
\cw{0, 0}\cb{10, 0}\cw{20, 0}\cb{30, 0}\cw{40, 0}
\put(13,-4){\vector(1,2){6}}
\end{picture}
\qquad
\begin{picture}(55,40)(-10,-5)
\put(-10,-3){\makebox(0,0)[cc]{(c)}}
\cb{0,30}\cw{10,30}\cb{20,30}\cw{30,30}\cb{40,30}
\cw{0,20}\cb{10,20}\cw{20,20}\cb{30,20}\cw{40,20}
\cb{0,10}\cb{10,10}\cw{20,10}\cb{30,10}\cb{40,10}
\cw{0, 0}\cb{10, 0}\cw{20, 0}\cb{30, 0}\cw{40, 0}
\put(13,-4){\vector(1,2){6}}
\end{picture}
\caption{Essential configurations for excess particle monomer: (a) excess particle monomer, (b,c) transient states. Centering is indicated by arrow.}
\label{fig_confE}
\end{figure}

The essential configurations at $q^2$ level include: empty c(2$\times$2) lattice, excess particle monomer (see Fig.~\ref{fig_confE}(a)), vacancy monomer, and transient states for both types of monomers (see Fig.~\ref{fig_confE}(b,c) for excess particle). Other transient states may be ignored. Because the empty lattice and the two kinds of monomers have different topological charges, monomers move independently at this level of accuracy. Let us now derive the kinetic equation for excess particle monomers. The approach may be generalized to any dimension. Here we consider $d$-dimensional case that makes formulae more obvious. The level $q^2$ should be replaced by $q^d$.

Let us denote the probability of the configuration in Fig.~\ref{fig_confE}(a) by $p_{\vec{\alpha}}$, where $\vec{\alpha}$ is a site indicated by the arrow in the figure. The transient states in Fig.~\ref{fig_confE}(b,c) are denoted by $p_{\vec{\sigma}}^{\vec{e}\vec{e}'}$. They are formed by vacancy at site $\vec{\sigma}$ and two excess particles at sites $\vec{\sigma}+\vec{e}$ and
$\vec{\sigma}+\vec{e}'$ (Figs.~\ref{fig_confE}(b) and \ref{fig_confE}(c) correspond to
$\vec{e}'\neq \vec{e}$ and $\vec{e}'=-\vec{e}$, respectively). Now the kinetic equation for excess particle monomer is given by
\begin{equation}  \label{eq_Emain}
\left\{
\begin{array}{l}
\dot{p}_{\vec{\alpha}} = -2d(2d-1)q^{-1}p_{\vec{\alpha}}+q^{1-2d}\sum_{\vec{e}\neq \vec{e}'}p_{\vec{\alpha}-\vec{e}}^{\vec{e}\vec{e}'},  \\
\dot{p}_{\vec{\sigma}}^{\vec{e}\vec{e}'} =
-2q^{1-2d}p_{\vec{\sigma}}^{\vec{e}\vec{e}'}+q^{-1}(p_{\vec{\sigma}+\vec{e}}+p_{\vec{\sigma}+\vec{e}'}),
\quad \vec{e}'\neq \vec{e}.
\end{array}
\right.
\end{equation}
In the first equation of this set the first rhs-term describes (a)$\rightarrow$(b,c) processes. Here $2d$ is the number of outer particles in monomer and $2d-1$ is the number of sites to which they can jump with the rate $q^{-1}$. The second rhs-term corresponds to the reverse processes which all have the same rate $q^{1-2d}$. In the second equation of the set the first rhs-term describes (b)$\rightarrow$(a) (if $\vec{e}+\vec{e}'\neq 0$) or (c)$\rightarrow$(a) (if $\vec{e}+\vec{e}'=0$) processes. Here the factor of 2 accounts for two possible particles which can jump to the site $\vec{\sigma}$. The last term corresponds to the reverse processes.

To obtain the correlation function we have to solve this set of equations with respect to $p_{\vec{\alpha}}$ up to the order $O(q^d)$. To do this we introduce a new variable
\begin{equation}
p'_{\vec{\alpha}}=\sum_{\vec{e}\neq \vec{e}'}p_{\vec{\alpha}-\vec{e}}^{\vec{e}\vec{e}'}
\end{equation}
and also a deformed Laplace operator (in continuous limit it reduces to standard laplacian)
\begin{equation}
\tilde{\Delta}_\vec{x}\varphi
=\frac{1}{4d}\sum_{\vec{e}\vec{e}'}\varphi_{\vec{x}+\vec{e}+\vec{e}'}-d\varphi_\vec{x}.
\end{equation}
Now (\ref{eq_Emain}) reduces to two compact equations:
\begin{equation}  \label{eq_E}
\left\{
\begin{array}{l}
\dot{p}_{\vec{\alpha}}=-2d(2d-1)q^{-1}p_{\vec{\alpha}}+q^{1-2d}p'_{\vec{\alpha}}, \\
\dot{p}'_{\vec{\alpha}}=-2q^{1-2d}p'_{\vec{\alpha}}+4d(2d-1)q^{-1}p_{\vec{\alpha}}+4dq^{-1}\tilde{\Delta}_{\vec{\alpha}}p.
\end{array}
\right.
\end{equation}
They can be solved by means of discrete $d$-dimensional Fourier transform: $f_\vec{x}\to\hat{f}(\vec{k})=\sum_\vec{x}f_\vec{x}e^{i\vec{k}\vec{x}}$. Its inverse is given by
\begin{equation}  \label{inv_Fourier}
f_\vec{x}=\frac{1}{(2\pi)^d}\int\limits_{[-\pi,\pi]^d}\hat{f}(\vec{k})e^{-i\vec{k}\vec{x}}\mathrm{d}\vec{k}.
\end{equation}
In the $k$-domain (\ref{eq_E}) transforms into two ordinary linear differential equations of the first order:
\begin{equation}  \label{eq_Efinal}
\left\{
\begin{array}{l}
\dot{\hat{p}}=-2d(2d-1)q^{-1}\hat{p}+q^{1-2d}\hat{p}',  \\
\dot{\hat{p}}'=-2q^{1-2d}\hat{p}'+4d(2d-1)q^{-1}\hat{p}-4q^{-1}\delta(\vec{k})\hat{p},
\end{array}
\right.
\end{equation}
where
\begin{equation}
\delta(\vec{k})=d^2-\left(\sum_{i=1}^{d}\cos k_i\right)^2.
\end{equation}
Their solution is expressed via two exponents $e^{\lambda_{1,2}t}$, where $\lambda_{1,2}$ are solutions of the equation
\begin{equation}
\lambda^2+2\lambda\left[d(2d-1)q^{-1}+q^{1-2d}\right]+4q^{-2d}\delta(\vec{k})=0.
\end{equation}
At small $q$ these solutions are given by $\lambda_1\approx-2q^{1-2d}$, describing nonpropagating short-time fluctuations, and $\lambda_2\approx-2q^{-1}\delta(\vec{k})$, corresponding to random displacements of a monomer as a whole.

Similarly, we may perform a Laplace transform of (\ref{eq_E}) with respect to variable $t$. In such a way Green's function of Eq.~\ref{eq_E} corresponding to $p_\vec{\alpha}$ (that means $p_\vec{\alpha}(t) =\sum_{\vec{\beta}}p_{\vec{\beta}}(0)G^\text{e}_{\vec{\alpha}-\vec{\beta}}(t)$) reduces to
\begin{multline}  \label{G_exact}
\tilde{G}^\text{e}_{\vec{\alpha}-\vec{\beta}}(s) =\frac{q}{2d}\left(1+\frac{sq^{2d}}{4d}\right) \\
\times\tilde{g}_{\vec{\alpha}-\vec{\beta}}\left(\frac{sq}{2d}\left[1+\frac{sq^{2d}}{4d}+d(2d-1)q^{2d-2}\right]\right),
\end{multline}
where $s$ is the Laplace variable, tilde marks transformed function, upper index ``e'' means ``the excess particle monomer'', and $g$ denotes Green's function of the equation
$\dot{\varphi}_\vec{x}=\tilde{\Delta}_\vec{x}\varphi$, which in $k$-domain reduces to
\begin{equation}  \label{g}
\hat{g}(t,\vec{k})=\exp\left\{\frac{t}{d}\left[\left(\sum_{i=1}^{d}\cos
k_i\right)^2-d^2\right]\right\}.
\end{equation}
Simple estimate of (\ref{G_exact}) shows that $G^\text{e}_{\vec{\alpha}-\vec{\beta}}(t) =g_{\vec{\alpha}-\vec{\beta}}\left(2dtq^{-1}\right)+O(q^{2d-2})$.

The kinetic equation for vacancy monomers is similar to that for excess particles but with the only difference that time $t$ is rescaled by the factor $q$. This means that $q^{-1}$ and $q^{1-2d}$ in (\ref{eq_E}) should be changed by 1 and $q^{2-2d}$, respectively. Therefore $G^\text{v}_{\vec{\sigma}-\vec{\tau}}(t) =g_{\vec{\sigma}-\vec{\tau}}(2dt)+O(q^{2d-2})$. The diffusion coefficients of monomers, derived from the mean square displacement $-\Delta_kg(t,k)|_{k=0}=2dDt$, are equal to $2dq^{-1}$ and $2d$ for excess particles and vacancies, respectively.

Summarizing, in the two-dimensional case at level $q^3$ the evolution of monomers is governed by the following Green's functions:
\begin{equation}  \label{Ge,Gv}
G^\text{e}_{\vec{\alpha}-\vec{\beta}}(t) =g_{\vec{\alpha}-\vec{\beta}}\left(4tq^{-1}\right), \quad
G^\text{v}_{\vec{\sigma}-\vec{\tau}}(t) =g_{\vec{\sigma}-\vec{\tau}}(4t),
\end{equation}
where $g$ is given by its Fourier transform (\ref{g}) with $d=2$. Note that $g_x$ is nonzero only if $|x_1+\ldots+x_d|$ is even. This simply means that monomers move only on their own sublattices.

For $|\vec{\alpha}-\vec{\beta}|\gg 1$ and $|\vec{\sigma}-\vec{\tau}|\gg 1$ we come to the diffusion equation used in our previous studies \cite{Argyrakis05,Chumak1,Chumak2}.

%subsec{Kinetic equation for flip-flop pairs}

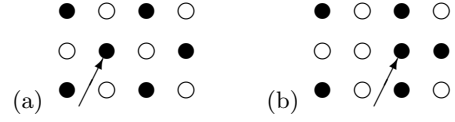
\begin{figure}[h]
\unitlength=1.5pt
\centering
\begin{picture}(50,30)(-10,-5)
\put(-10,-3){\makebox(0,0)[cc]{(a)}}
\cb{0,20}\cw{10,20}\cb{20,20}\cw{30,20}
\cw{0,10}\cb{10,10}\cw{20,10}\cb{30,10}
\cb{0, 0}\cw{10, 0}\cb{20, 0}\cw{30, 0}
\put(3,-4){\vector(1,2){6}}
\end{picture}
\qquad
\begin{picture}(50,30)(-10,-5)
\put(-10,-3){\makebox(0,0)[cc]{(b)}}
\cb{0,20}\cw{10,20}\cb{20,20}\cw{30,20}
\cw{0,10}\cw{10,10}\cb{20,10}\cb{30,10}
\cb{0, 0}\cw{10, 0}\cb{20, 0}\cw{30, 0}
\put(13,-4){\vector(1,2){6}}
\end{picture}
\caption{Essential configurations for flip-flop pairs. Centering is indicated by arrow.}
\label{fig_confFF}
\end{figure}

Essential configurations for flip-flop pairs include only empty c(2$\times$2) lattice, see Fig.~\ref{fig_confFF}(a), and flip-flop pair, see Fig.~\ref{fig_confFF}(b). We denote the probability of these configurations as $p_0$ and $p_\vec{e}$, respectively. The kinetic equations are simple:
\begin{equation}  \label{eq_FFmain}
\left\{
\begin{array}{l}
\dot{p}_0 = -2dp_0+2dq^{1-2d}p_\vec{e},  \\
\dot{p}_\vec{e} = -q^{1-2d}p_\vec{e}+p_0.
\end{array}
\right.
\end{equation}
Here $-2dp_0$ term describes (a)$\rightarrow$(b) processes, where $2d$ is the number of nearest neighbor sites to which the particle may jump.

The corresponding Green's functions are given by
\begin{gather}
G^{\text{ff}}_{\vec{0}\vec{0}}=\frac{\nu+e^{-(2d+\nu)t}}{2d+\nu}, \qquad
G^{\text{ff}}_{\vec{0}\vec{e}}=\frac{1-e^{-(2d+\nu)t}}{2d+\nu},  \notag \\
G^{\text{ff}}_{\vec{e}\vec{0}}=\nu G^{\text{ff}}_{\vec{0}\vec{e}}, \quad
G^{\text{ff}}_{\vec{e}\vec{e}}=\frac{1}{2d}\frac{2d+\nu
e^{-(2d+\nu)t}}{2d+\nu}+\frac{2d-1}{2d}e^{-\nu t},  \notag \\
G^{\text{ff}}_{\vec{e}\vec{e}'}=\frac{1}{2d}\frac{2d+\nu
e^{-(2d+\nu)t}}{2d+\nu}-\frac{1}{2d}e^{-\nu t},\; \vec{e}\neq \vec{e}',  \label{Gff}
\end{gather}
where $\nu=q^{1-2d}$. It follows from Eq.~\ref{eq_FFmain} that in the case of $q\ll 1$ the relaxation time of the flip-flop fluctuations is given by $q^{2d-1}$.

\section{Correlation function} \label{sec_cf}

Now we can obtain the correlation function. There is no correlation in the motion of the two types of monomers and flip-flop pairs. They evolve independently and give additive contributions to the total correlation function. To find these contributions we can use the general formula for the correlation function of independent random walks, derived below.

%subsec{Correlation function for random walker}

Let $N$ particles randomly walk on a lattice. Denote by $\xi_i$ the position of $i$-th particle. The number of particles in site $\vec{x}$ is given by
\begin{equation}
n_\vec{x}(t)=\sum_{i=1}^{N}\mathcal{I}\{\xi_i(t)=\vec{x}\},
\end{equation}
where $\mathcal{I}\{A\}$ is the indicator of an event $A$. In what follows we take into account the identity of particles. In this case their average number at a given site, that is the average on possible trajectories with fixed initial distribution, reduces to
\begin{equation}
\left\langle n_\vec{x}(t)\right\rangle =N\left\langle\mathcal{I}\{\xi(t)=\vec{x}\}\right\rangle
=N\mathrm{P}\{\xi(t)=\vec{x}\}=Np_\vec{x}(t),
\end{equation}
where $\xi$ is the position of any selected particle and the formula $\left\langle\mathcal{I}\{A\}\right\rangle=\mathrm{P}\{A\}$ is used. The correlation function can be calculated as follows ($t\geq s$):
\begin{equation}
\begin{split}
&\left\langle n_\vec{x}(t)n_\vec{y}(s)\right\rangle
=\sum_{i}\left\langle\mathcal{I}\{\xi_i(t)=\vec{x},\xi_i(s)=\vec{y}\}\right\rangle \\
&\qquad +\sum_{i\neq j} \left\langle\mathcal{I}\{\xi_i(t)=\vec{x},\xi_j(s)=\vec{y}\}\right\rangle \\
&\quad =N\left\langle\mathcal{I}\{\xi(t)=\vec{x},\xi(s)=\vec{y}\}\right\rangle \\
&\qquad +N(N-1)\left\langle\mathcal{I}\{\xi(t)=\vec{x}\}\right\rangle \left\langle\mathcal{I}\{\xi(s)=\vec{y}\}\right\rangle \\
&\quad =Np_{\vec{y}\vec{x}}(s,t)+N(N-1)p_\vec{x}(t)p_\vec{y}(s).
\end{split}
\end{equation}
Hence the correlation function will be given by
\begin{multline}  \label{cf2}
\left\langle\delta n_\vec{x}(t)\delta n_\vec{y}(s)\right\rangle =\left\langle
n_\vec{x}(t)n_\vec{y}(s)\right\rangle -\left\langle n_\vec{x}(t)\right\rangle\left\langle n_\vec{y}(s)\right\rangle \\
=Np_\vec{y}(s)[G_{\vec{y}\vec{x}}(t-s)-p_\vec{x}(t)].
\end{multline}
For homogeneous lattice at equilibrium (\ref{cf2}) reduces to
\begin{equation}  \label{cf_hom}
\left\langle\delta n_\vec{x}(t)\delta n_\vec{y}(s)\right\rangle
=n[G_{\vec{x}-\vec{y}}(t-s)-G_{\vec{x}-\vec{y}}(\infty)],
\end{equation}
where $n=\left\langle n_\vec{x}(\infty)\right\rangle$ is the equilibrium concentration of particles.

%subsec{Correlation function for monomers}

Straightforward application of (\ref{cf_hom}) to monomers gives their contribution to the correlation function to be
\begin{equation}  \label{cf_EV}
\begin{split}
&\left\langle\delta n_\vec{\alpha}(t)\delta n_\vec{\beta}(0)\right\rangle^\text{e}
=n^\text{e}G^\text{e}_{\vec{\alpha}-\vec{\beta}}(t)
=n^\text{e}g_{\vec{\alpha}-\vec{\beta}}(4tq^{-1}),   \\
&\left\langle\delta n_\vec{\sigma}(t)\delta n_\vec{\tau}(0)\right\rangle^\text{v}
=n^\text{v}G^\text{v}_{\vec{\sigma}-\vec{\tau}}(t) =n^\text{v}g_{\vec{\alpha}-\vec{\beta}}(4t),
\end{split}
\end{equation}
where the concentrations of monomers were calculated in \cite{Chumak1}:
\begin{equation}
n^\text{e,v}\approx\sqrt{\left(c-\frac{1}{2}\right)^2+q^4}\pm\left(c-\frac{1}{2}\right)
\end{equation}
(upper sign is for excess particle monomers). Other components like $\left\langle\delta n_\vec{\alpha}\delta n_\vec{\sigma}\right\rangle$ are zero for monomers in the approximation explained above.

%subsec{Correlation function for flip-flop pairs}

For flip-flop pairs the factor $n$ in (\ref{cf_hom}) is unity (it is the concentration of configurations in Fig.~\ref{fig_confFF}(a)). We should take also into account that each particle in the almost empty sublattice may appear there due to a jump from any of its nearest neighbor sites. In this way we obtain the following nonzero components of the correlation function:
\begin{equation}  \label{cf_FF}
\begin{split}
\langle\delta n_\vec{x}(t)\delta n_\vec{x}(0)\rangle^\text{ff} & \approx 4q^3e^{-q^{-3}t},  \\
\langle\delta n_\vec{x}(t)\delta n_{\vec{x}+\vec{e}}(0)\rangle^\text{ff} & \approx -q^3e^{-q^{-3}t}.
\end{split}
\end{equation}

\section{Fluctuations in probe area: comparison with MC simulations} \label{sec_MC}

%subsec{The relation between correlation function and fluctuations in probe area}

Now according to (\ref{dN2}) we have to sum up the derived two-point correlation functions over the square probe area of size $L\times L$ to obtain the quantity $S_L(t)$, which we can compare with the results of MC simulations.

From (\ref{cf_EV}) the contribution of monomers will be given by
\begin{equation}  \label{dN2_EV}
S_L^\text{ev}(t) =\frac{n^\text{e}}{L^2}\sum_{\vec{\alpha},\vec{\beta}} g_{\vec{\alpha}-\vec{\beta}}\left(4tq^{-1}\right) +\frac{n^\text{v}}{L^2}\sum_{\vec{\sigma},\vec{\tau}} g_{\vec{\sigma}-\vec{\tau}}(4t).
\end{equation}
The sum in (\ref{dN2_EV}) can be evaluated by using the identity
\begin{equation}  \label{sum_xy}
\sum_{\vec{x},\vec{y}}f_{\vec{x}-\vec{y}}
=\frac{1}{(2\pi)^d}\int\limits_{[-\pi,\pi]^d}\hat{f}(\vec{k})
\left|\sum_\vec{x}e^{-i\vec{k}\vec{x}}\right|^2\mathrm{d}\vec{k},
\end{equation}
which takes place if $\vec{x}$ and $\vec{y}$ vary within the same domain. In particular, for parallelepiped ($x_i=\overline{0,L_i-1}$)
\begin{equation}
\left|\sum_\vec{x}e^{-i\vec{k}\vec{x}}\right|
=\prod_{i=1}^d\frac{\sin\frac{k_i L_i}{2}}{\sin\frac{k_i}{2}}.
\end{equation}
In case of even $L$ the sums in (\ref{dN2_EV}) and (\ref{sum_xy}) are connected by the identity $\sum_{\vec{\alpha},\vec{\beta}}=\frac{1}{2}\sum_{\vec{x},\vec{y}}$.

For flip-flop pairs we can obtain explicit expression by using the formula (time arguments are omitted)
\begin{equation}
\begin{split}
S_L^\text{ff} &=\left\langle\delta n_{(00)}\delta n_{(00)}\right\rangle \\
&+\frac{4}{L}\sum_{m=1}^{L-1}(L-m)\left\langle\delta n_{(m0)}\delta n_{(00)}\right\rangle  \\
&+\frac{4}{L^2}\sum_{m=1}^{L-1}(L-m)^2\left\langle\delta n_{(mm)}\delta n_{(00)}\right\rangle \\
&+\frac{8}{L^2}\sum_{m=2}^{L-1}\sum_{l=1}^{m-1}(L-m)(L-l)\left\langle\delta n_{(ml)}\delta n_{(00)}\right\rangle.
\end{split}
\end{equation}
Thus, from (\ref{cf_FF}) it follows
\begin{equation}  \label{dN2_FF}
S_L^\text{ff}(t)=\frac{4}{L}q^3\exp(-q^{-3}t).
\end{equation}
This expression does not depend on $c$ at level $q^3$. The total correlation function is given by
\begin{equation}  \label{S_tot}
S_L=S_L^\text{ev}+S_L^\text{ff}.
\end{equation}

%subsec{The contribution of flip-flop jumps to fluctuations}

\begin{figure}
\centering
\includegraphics{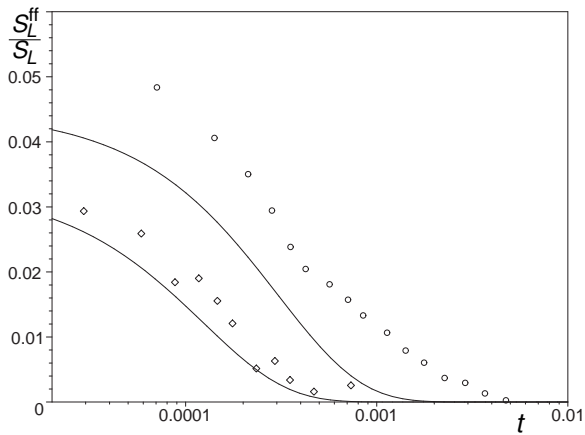}
\caption{Relative contribution of fluctuations due to flip-flop jumps for $\phi=2.7$ (top) and $3.0$ (bottom), $L=6$. Circles and diamonds are MC data, while full lines are theoretical calculations.} \label{fig_flucFF}
\end{figure}

\begin{figure}
\begin{minipage}[b]{1.0\linewidth}
\centering
\includegraphics{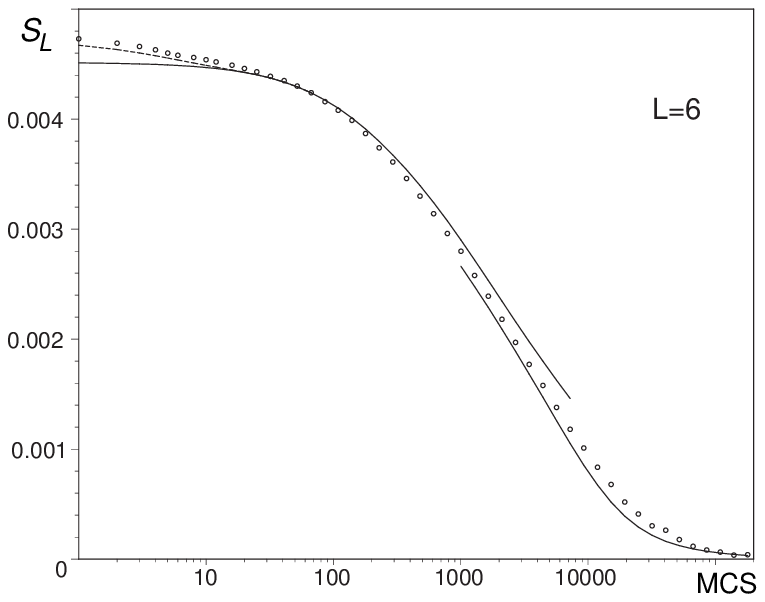}
\end{minipage}
\begin{minipage}[b]{1.0\linewidth}
\centering
\includegraphics{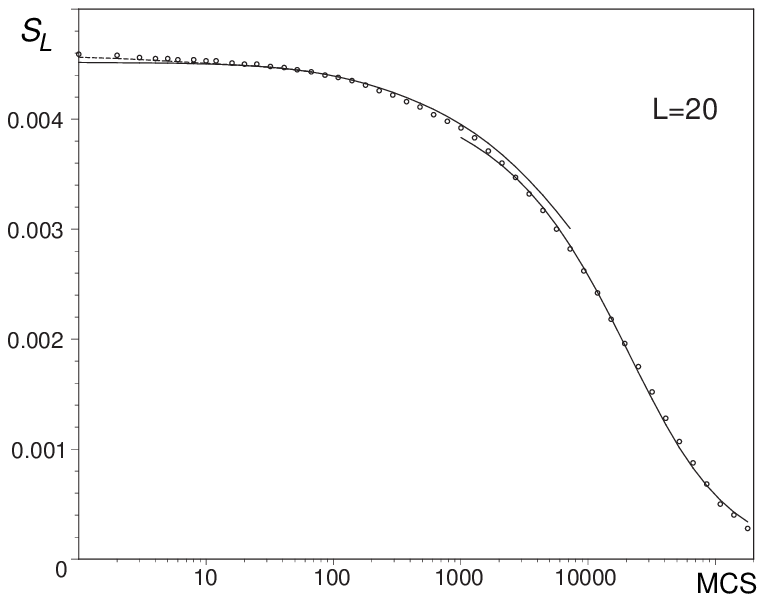}
\end{minipage}
\caption{Correlation function of fluctuations in probe area per site vs. MC steps for $\phi=2.70$, $c=0.5$, $L$ is indicated. 1~MCS corresponds to time $[4(1+e^{\phi})(1+e^{2\phi})]^{-1}$. Circles are MC data. The solid line at short times is the correlation function of fluctuations due to random walk of monomers only given by Eq.~\ref{dN2_EV}. The dash line (merging with solid at 20 MCs) is the same quantity with flip-flop jumps accounted as well (Eq.~\ref{S_tot}). The solid line at long times is the correlation function of fluctuations from the diffusion and recombination of monomers from Ref.~\cite{Argyrakis05}.}
\label{fig_fluc}
\end{figure}

To analyse the contribution of flip-flop jumps to fluctuations it should be noted that $S_L^\text{ff}$ is a decreasing function of size $L$. This point may be easily understood by the observation that displacements of only border particles (whose total number is proportional to $L$) determine fluctuations of $\delta N$. Relative contribution of flip-flop jumps to $S_L$ is shown in Fig.~\ref{fig_flucFF}, from which it follows that this contribution is negligibly small when $\phi\gtrsim 2.5$ (see also Fig.~\ref{fig_fluc}). This result agrees well with the calculations shown in Fig.~\ref{fig_cfat0}. Having zero topological charge, flip-flop pairs contribute insignificantly to the correlation function at times of the order or longer than $q^3$. On the other hand, Fig.~\ref{fig_flucFF} shows the adequacy and good accuracy of our approach. The main shortcoming of (\ref{dN2_FF}), that becomes apparent for small $\phi$ in Fig.~\ref{fig_flucFF}, originates from the crude underestimation of concentration of flip-flop pairs in (\ref{n_ff}).

%subsec{Validation of the developed theory}

For large values of interaction parameter ($\phi\gtrsim 2.5$) the motion of monomers gives dominant contribution. It was shown in \cite{Argyrakis05} that in the case of large times the generation-recombination processes of monomers should also be taken into account. These processes are responsible for establishing local equilibrium in the defect system and determine the dissipation of smooth spatio-temporal inhomogeneities of the particle density and their fluctuations. Bringing together the approach of the present paper, which provides adequate description of the system at kinetic scales (the characteristic length is of the order of lattice constant and the characteristic time is of the order of defect living-time in a given site), with that developed in \cite{Argyrakis05} for hydrodynamic scales we get a description of fluctuations at any time, both short and long. Thus, it becomes possible to compare theoretical data with the results of MC simulations in the whole range of computer simulations, Fig.~\ref{fig_fluc}. A good agreement between the two can be seen.

\section{Discussions}

A very simple physical picture explaining the nature of dynamic correlations in the ordered c(2$\times$2) phase may be outlined from this analysis. In the case of sufficiently strong particle-particle interaction ($\phi\gtrsim 2.5$ or $T\lesssim 0.7\, T_c$) and half-filling ($|c-1/2|\ll q$) the dominant contribution to the dynamic correlation function is due to random walks of two types of structural defects: excess particle monomers and vacancy monomers, which have concentrations $q^2$. At the same time, another two processes must be accounted for accurate description of the correlation function: 1) at short times -- additional fluctuations caused by flip-flop pairs with concentration $2q^3$; 2) at long times -- faster decay of correlations due to generation-recombination of monomers, $O(q^4)$. If $|c-1/2|\gg q$ then one of two types of monomers is in majority with concentration $|2c-1|$, and its random walking determines the correlation function. In this case dimers with concentration $|2c-1|^2$ must also be accounted, which decrease the collective diffusion coefficient.

The decay of correlations is described by several characteristic times. First of all, it is the living time of flip-flop pair (Fig.~\ref{fig_confFF}(b)), which is of the order of $q^3$. Two other characteristic times are connected with the duration of monomer jumps ($q^3$ for excess particles and $q^2$ for vacancies). They are equal to living times of transient configurations (Fig.~\ref{fig_confE}(b,c)) formed in the course of two-step defect displacements. The relative contribution of these transient configurations is of the order of $q^4$. At this time scale all other short-living configurations snapshoted in Fig.~\ref{fig_snapshot2} also decay. At moderate times only monomers (and their groups) are essential. The corresponding characteristic times are $q$ for excess particle monomers and 1 for vacancy monomers, which are time intervals between their successive jumps. Random walk of monomers results in power-law decay of correlations, $t^{-1}$ as $t\to\infty$. At very long times generation-recombination processes become essential. They speed up the decay of correlations though the asymptotic behavior of the correlation function is still $t^{-1}$. This is in full accordance with some rigorous lower estimates (\cite{Bernardin05} and ref. therein) and with long-time asymptotics of relaxation of concentration fluctuations in $A+B \leftrightarrow C$ reversible diffusion-limited reaction considered in \cite{Burlatsky89} (see also \cite{benAvraham00,Oshanin96} for details).

Summarizing, the proposed method of essential configurations makes it possible to describe the evolution of structural defects and to obtain the dynamic correlation function for a lattice gas with nearest neighbor repulsion in the ordered c(2$\times$2) phase. Our calculations explain MC simulations reported in \cite{Argyrakis05} at short times and show the range of times where the diffusional approach developed in \cite{Argyrakis05,Chumak1,Chumak2} is not applicable.

Further development of the present work includes the extension of the method to $n^2$ level, that is necessary for accurate description of generation-recombination processes. The main complication is that we must proceed from one-particle description to a many-particle one.

Concluding, our consideration shows that for known particle-particle interaction and jump rate mechanisms, the correlation function of particle number fluctuations in small probe area can be calculated for the specific model of the ordered lattice gas. Hence the comparison of the analytical and experimental data concerning short time correlations becomes possible. In principle, the information about individual particle jumps or complex jumps accompanied with the defect displacements may be extracted from such comparison. This paper may be considered as the attempt to get better understanding of short-time correlations in ordered systems.

\begin{acknowledgments}
We thank Prof. P.~Holod for helpful discussions. This work was supported by the NATO
Collaborative Linkage Grant PST.CLG.979878.
\end{acknowledgments}

\end{document}